\documentstyle[11pt,epsfig]{article}
\newcommand{\bea}{\begin{eqnarray}}
\newcommand{\eea}{\end{eqnarray}}

\def\simlt{\stackrel{<}{{}_\sim}}
\def\simgt{\stackrel{>}{{}_\sim}}

\begin{document}
\begin{titlepage}

\thispagestyle{empty}

\vspace{0.2cm}

\title{The Constraints on CP Violating Phases in models with a dynamical
gluino phase
\author{M\"{u}ge Boz\\
Hacettepe University, Department of Physics,\\  
06532 Ankara, Turkey\\}}
\date{}
\maketitle

\begin{center}\begin{minipage}{5in}

\begin{center} ABSTRACT\end{center}
\baselineskip 0.2in
{We have analyzed the electric dipole moment and the Higgs mass constraints on the
supersymmetric  model which offers dynamical solutions to the $\mu$
and  strong CP problems. The  trilinear coupling phases, and
$\tan\beta-|\mu|$ are strongly
correlated,  particularly in the low-$\tan\beta$ regime. Certain values of the phases of the 
trilinear couplings are forbidden, whereas
the CP violating phase from the chargino sector 
is imprisoned to lie near a CP conserving point, by the Higgs mass and
electric dipole moment constraints. }
\end{minipage}
\end{center}
\end{titlepage}

\eject

\rm
\baselineskip=0.25in
\section{Introduction}
In the standart electroweak theory (SM) the single phase in the CKM 
matrix, $\delta_{CKM}$, is the unique source for both flavour and CP violations.
In the supersymmetric (SUSY) extensions of the standart model,
there exist novel sources for both flavour and CP violations coming from 
the soft supersymmetry breaking mass terms~\cite{Dugan}. 
The new sources of CP violation can be probed via the  flovor conserving processes such as the
electric dipole moments (EDMs)~\cite{Edm1,Edm2,Edm4,Edm5,IbrahimNath2,Ritz} 
of the particles, and the Higgs system~\cite{Pilaftsis1,Demir1,Choi1,Carena1,IbrahimNath1,Ham,Boz1},
leading to novel signatures at high-energy colliders~\cite{Carena2}. 
On the other hand, a searching platform for  flavour violation is the
Higgs-mediated  flavour changing processes~\cite{BozPak,Bll}. 

The SUSY CP problem is one of the main hierarchy problems that SUSY theories possess.
In fact, the EDMs of the  neutron and the electron, severely constrain the 
strength of the CP violation. To evade these constraints,
without suppressing the CP violating phases of the theory,
several works have been carried out in the existing literature
which include  choosing the SUSY CP phases very small ($\simlt
{\cal O}(10^{-3})$)~\cite{Edm1},  sparticle masses large~\cite{Edm2},
arranging for partial cancellations among the different 
contributions to the EDM~\cite{Edm4,Edm5}, and 
suppressing the phases only in the third generation~\cite{Voloshin,Chang,Pilaftsis3}
in the framework of the effective supersymmetry~\cite{GiudiceBinetruy}. 

Clearly, even if the SUSY CP problem is solved,
there are still other hierarchy problems in SUSY theories:
The strong CP problem whose source is the neutron electric dipole moment
exceeding the present bounds by nine orders of magnitude~\cite{Harris}, 
and the $\mu$ puzzle, concerning the Higgsino Dirac
mass parameter ($\mu$), which follows from the
superpotential of the model. A simultaneous solution to both those
problems have been shown to exist with a SUSY
version~\cite{DemirMa1} of the Peccei-Quinn mechanism,
by using a new kind of axion~\cite{Peccei,KimDine}, 
which couples to the gluino
rather than to quarks. 
In this model the invariance of the supersymmetric Lagrangian and all supersymmetry
breaking terms  under $U(1)_R$ is guaranteed  by  promoting the ordinary   
$\mu$ parameter to a composite operator involving the gauge singlet $\hat{S}$
with unit $R$ charge. When the scalar component of the singlet develops vacuum
expectation value (VEV) around the Peccei--Quinn scale $\sim 10^{11}~{GeV}$  an
effective $\mu$ parameter $\mu\sim \mbox{a TeV}$ is induced. 
Besides, the low energy theory is identical to minimal
SUSY  model with all sources of soft SUSY phases
except for the fact that the soft masses are all expressed in terms of
the $\mu$ parameter through appropriate flavour matrices~\cite{DemirMa1}. 
The effective Lagrangian of the theory possesses all sources of CP violation 
through  the  complex trilinear couplings ($A_{t,b,e}$),  
\begin{eqnarray}
\label{1}
A_{t}=\mu^{*} \ k_t,\,\,\,\, A_{b}=\mu^{*} \ k_b,\,\,\,\,A_{e}=\mu^{*} \ k_e~,
\end{eqnarray}
the effective $\mu$ parameter itself, and the  gaugino masses 
\begin{eqnarray}
M_{1}=k_{1}\ \mu^{*}~,\ \ \ M_{2}=k_{2}\ \mu^{*},~\ \ \ M_{3}=|k_{3}|\ \mu^{*}~,
\end{eqnarray}
where $k_{t(b,e)}$  and  $k_{1,2,3}$ are the dimensionless complex  parameters.

There are other parameters in the model, namely,  the 
squark and  the slepton soft masses which assume the form: 
\begin{eqnarray}
\label{2}
M_{\tilde{Q}}^{2}&=&k_{\tilde Q}^{2}\ |\mu|^{2}~,\ \ \ M_{\tilde{u}}^{2}=
k_{\tilde u}^{2}\ |\mu |^{2}~,\ \ \ M_{\tilde{d}}^{2}=
k_{\tilde d}^{2}\ |\mu |^{2}~,\nonumber\\
M_{\tilde{L}}^{2}&=&k_{\tilde L}^{2}\ |\mu|^{2}~,\ \ \
M_{\tilde{e}}^{2}=k_{\tilde e}^{2}\ |\mu|^{2}~,
\end{eqnarray} 
where $k_{\tilde L,\tilde e}$ are real parameters.
As suggested by Eqs.(1-3) all soft masses in the theory
are fixed in terms of the $\mu$ parameter. 
Therefore, by  naturalness, 
all dimensionless parameters ($|k_i|$) are expected to be of order $\sim
{\cal O} (1)$. 

The main purpose of this work is to study the
effects of EDM and Higgs mass constraints
on the CP violating phases of the  model. 
To be specific  we consider the electron EDM, and  analyze  
the  various parameter planes to determine the possible constraints on
the $\mu$ parameter and the physical phases of the model;
that is the phases of stop ($A_{t}$), sbottom ($A_{b}$), and selectron
($A_{e}$) tri-linear couplings:   
\begin{eqnarray}
\label{phasetb}
\varphi_{A_{t}}= Arg[\mu^{*}\ k_t]~, \,\,\,
\varphi_{A_{b}}= Arg[\mu^{*} \ k_b]~,\,\,\,
\varphi_{A_{e}}= Arg[\mu^{*} \ k_e]~, 
\end{eqnarray}
and,  of the gaugino masses: 
\begin{eqnarray}
\label{phasecn}
\varphi_{1}=Arg[\mu^{*} \ k_1]~, \,\,\,
\varphi_{2}= Arg[\mu^{*} \ k_2].~
\end{eqnarray}
After having determined the possible 
constraints on the CP violating  phases of the model, from the Higgs mass bound~\cite{LEP1,LEP2} 
and the eEDM, we will study  the dependence of 
the eEDM on the CP violating phases of the theory,
by considering various parameter planes.

The organization of the work is as follows:
In Section 2, we study the one and two loop  contributions to the 
eEDM for the model under concern.
In Section 3, we carry out  the numerical analysis, to study the
effects of the EDM and Higgs mass constraints on the 
CP violating phases of the theory. The results are summarized in Section 4.

\section{Electron Electric Dipole Moment}
The model under concern does not
match to the effective supersymmety, since all sparticles acquire similar
masses. Therefore,  in our analysis we take into account of  both one and two-loop
contributions to eEDM: 
\begin{eqnarray} 
\bigg(\frac{d_e}{e}\bigg)&=&\bigg(\frac{d_e}{e}\bigg)^{1-loop}+\bigg(\frac{d_e}{e}\bigg)^{2-loop}~.
\label{edm}
\end{eqnarray}
We would like to note that in our presentation we will 
follow the detailed works of Ibrahim and
Nath~\cite{Edm4}, and Pilaftsis~\cite{Pilaftsis3}.
However,  we differ in the sense that in our analysis all the chosen parameters are specific to the 
gluino-axion model, namely all the soft mass parameters 
in this theory are fixed in terms of the dynamically--generated $\mu$
parameter,  whereas the dimensionless parameters involved
are naturally of order ${\cal{O}}(1)$ and source the SUSY CP violation.

The main  contributions to the one-loop eEDM come from the  neutralino, and
chargino exchanges, and can be calculated as~\cite{Edm4}: 
\begin{eqnarray}
\label{neutralinochargino}
\bigg(\frac{d_e}{e}\bigg)^{1-loop}&=& \frac{\alpha } {4 \pi
 s^2_{W}}\Bigg\{
\sum_{k=1}^{2} \sum_{i=1}^4 {  \mbox{Im} [{\eta}_ {e_{ik}}] \frac { m_{{\chi}_{i}^{0}}} 
{ m_{ {\tilde e}_{k}}^2} 
B\bigg(\frac  { m^{2}_{{\chi}_{i}^{0}}}   
{m_{ {\tilde e}_{k}}^2} \bigg)}\nonumber\\
&+& 
\frac {m_e}{\sqrt{2} \ c_\beta \  m_W \ m_{{\tilde \nu}_{e}}^2}
\sum_{i=1}^2 { m_{  {\chi}_{i}^{+}}   \mbox{Im} [U_{i2}^{*} V_{i1}^{*}]
A\bigg(\frac  {m^{2}_{{\chi}_{i}^{+}}}   
{m_{ {\tilde \nu}_{e}}^2}\bigg)}\Big\}~,
\end{eqnarray}
where
\begin{eqnarray}
{\eta}_{e_{ik}}&=&- \bigg[ 
\bigg (t_W {\cal{N}}_{1i}+ {\cal{N}}_{2i}  \bigg) \tilde{
{\cal{S}}}_{e1k }^{*}
+ \frac{m_e }{m_W c_{\beta} } {\cal{N}}_{3i} \tilde{ {\cal{S}}}_{e2k }^{*}
\bigg]\nonumber\\
&\times&
\bigg[t_W {\cal{N}}_{1i}  \tilde{ {\cal{S}}}_{e2k}
+\frac{m_e}{2 m_W c_{\beta}} {\cal{N}}_{3i}  \tilde{ {\cal{S}}}_{e1k}
\bigg]~,
\end{eqnarray}
and we set for convenience, $s_{W}(c_{W})=\sin{\theta_{W}}(\cos{\theta_{W}})$, \,
$t_{W}=\tan{\theta_{W}}$. In our analysis,
the squark  $(\mbox{mass})^2$ matrices,  expressed in terms of the parameters of the model,
\begin{eqnarray}
\label{sfermionmat}
\ \left( \begin{array}{cc} k_{{\tilde f}_{L}}^{2}
|\mu|^{2} + 
m^2_f + c_{ 2 \beta}  
 (T_{3f}- Q_f s_{W}^2 ) M_Z^2  
& m_f \,  \mu  (k_f -   R_{f})\\  
m_f  \,  \mu^{*} (k_f -  R_{f}) &
k_{{\tilde f}_{R}}^{2}  |\mu|^{2}  + 
m^2_f + c_{ 2 \beta} Q_f  M_Z^2    s_{W}^2  \end{array} \right)~,
\end{eqnarray}
can be  diagonalized via the unitary rotation: 
\begin{eqnarray}
\label{bb}
\tilde{ {\cal{S}}}_{f }^{\dagger}\, \widetilde{M}^2_f\,
\tilde{{\cal{S}}}_{f} =
\mbox{diag}\left(m^{2}_{\tilde{f}_1},
m^{2}_{\tilde{f}_2}\right)~.
\end{eqnarray}
where  $R_f= ({t_{\beta}}^{2 T_{3f}})^{-1}$,  and we set
$t_{\beta}$=$\tan\beta$, $c_{ 2 \beta}$=$\cos 2 \beta$.
Therefore, the eigenstates $({\tilde e}_{1}, {\tilde
e}_{2})$ in  (\ref{neutralinochargino}) can be obtained in analogy with (\ref{bb}) 
for $f=e$ ($Q_e= -1$, $T_{3e}=-\frac{1}{2}$).
Similar analysis can be performed for the sneutrinos, using 
\begin{eqnarray}
\label{sneutrinomat}
\widetilde{M}^2_{\nu_{e}} =
\ \left( \begin{array}{cc} k_{ {\tilde e} _{L}}^{2}
|\mu|^{2} +  c_{ 2 \beta}  
T_{3 \nu_{e}}  M_Z^2  
& 0\\  
0 &
M_G^2 \end{array} \right)~,
\end{eqnarray}
where $T_{3\nu _{e}}= \frac{1}{2}$, and  $M_G$ is the right-handed sneutrino
mass. Moreover, the neutralino and chargino masses in Eq. (\ref{neutralinochargino}) can be obtained by the following
transformations:
\begin{eqnarray}
{\cal{N}}^{T} \,  M_{{\chi}^{0}} \,  {\cal{N}} =\mbox{diag}\left(m_{\chi^{0}_1}, \cdots,
m_{\chi^{0}_4}\right)~,
\label{neutmat}
\end{eqnarray}
\begin{eqnarray}
\label{def}
{\cal{U}}^{*} M_{C} {{\cal{V}}^{-1}} = \mbox{diag} (m_{ {\chi}_{1}^{+} }, \, m_{ {\chi}_{2}^{+}})~.
\end{eqnarray}

For the two-loop eEDM, the dominant effects  
originate from the couplings of the Higgs bosons to stop-sbottom quarks,
top-bottom quarks and charginos \cite{Pilaftsis3}.  
The Higgs bosons couplings to stop-sbottom quarks
can be calculated as \cite{Pilaftsis3}: 
\begin{eqnarray}
\label{stopsbottom}
\bigg(\frac {d_e}{e}\bigg)^{2-loop}_{H_i-\tilde q}&=& 
a_0 \alpha m_e 
\sum_{i=1}^3 \frac{- t_\beta {\cal R}_{i3}}{m_{h_{i}}^2}   
\Bigg\{ \frac{4}{9}  \frac{2  m_{t}^2}{v^2 \Delta_{ {\tilde t} }} 
\bigg[\frac{ \mbox{Im} [ {\cal Z}_{1 t}]
{\cal R}_{i3}}{s_{\beta}^2}-\frac{ \mbox{Re} [ {\cal Z}_{2 t}]
{\cal R}_{i1}}{s_{\beta}}\nonumber\\
&+&\frac{\mbox{Re} [{\cal Z}_{3 t} ]
{\cal R}_{i2}}{s_{\beta}} \bigg]{\cal F}_{\tilde t}^{-}
-\frac{4}{9}\frac{2  m_{t}^2}{v^2}\frac{{\cal R}_{2i}}{s_{\beta}}
{\cal F}_{\tilde t}^{+} -\frac{1}{9} \frac{2  m_{b}^2}{v^2}\frac{{\cal R}_{1i}}{c_{\beta}}
{\cal F}_{\tilde b}^{+} \nonumber\\
&+&  \frac{1}{9}  
 \frac{2  m_{b}^2}{v^2 \Delta_{ {\tilde t} }} 
\bigg[\frac{ \mbox{Im} [ {\cal Z}_{1 b}]
{\cal R}_{i3}}{c_{\beta}^2}-\frac{ \mbox{Re} [ {\cal Z}_{2 b}]
{\cal R}_{i2}}{c_{\beta}}
+\frac{\mbox{Re} [{\cal Z}_{3 b} ]
{\cal R}_{i1}}{c_{\beta}} \bigg]{\cal F}_{\tilde b}^{-}
\Big \}~,
\end{eqnarray}
where $a_o= 3/ 32 \pi^3$,\,\, and 
$\Delta_{ {\tilde q}}$ is the squark splitting $(m_{{\tilde
q}_{2}}^2-m_{{\tilde q}_{1}}^2)$.
Here,  ${\cal F}_ q^ {\pm}=F\big(m_{{\tilde q}_{1}}^2 , M_A^2 \big) \pm
F\big(m_{{\tilde q}_{2}}^2, M_A^2 \big)$ are the loop functions~\cite{Pilaftsis3}.    

In Eq. (\ref{stopsbottom}), we define 
\begin{eqnarray}
{\cal Z}_{1 t(b)}= k_{t(b)}~,\,\,\,\,
{\cal Z}_{1 t(b)}= k_{t(b)}-t_{\beta}^{-1}(t_{\beta})~,\,\,\,\,{\cal Z}_{2  t(b)}=
|k_{t(b)}|^{2}|\mu|^2-k_{t(b)}^{*} t_{\beta}^{-1}(t_{\beta})~,
\end{eqnarray}
where  $k_{t(b)}$
are given by Eq. (\ref{phasetb}).

Moreover, the radiatively corrected Higgs masses ($m_{h_{i}}$) in Eq.
(\ref{stopsbottom}) can be obtained by the 
diagonalization of the  Higgs mass--squared matrix  by the similarity 
transformation: 
\begin{eqnarray}
{\cal{R}}M_{H}^{2}{\cal{R}}^{T}= {\rm diag}(m_{h_{1}}^{2},
m_{h_{2}}^{2}, m_{h_{3}}^{2})~,
\label{diagonal}
\end{eqnarray}
where ${\cal{R}}{\cal{R}}^{T}=1$. In our analysis, we define $h_3$ to be the
lightest of all three Higgs bosons, and $\rho_3$ to be its percentage  CP
component
($\rho_{3}=100\times |{\cal{R}}_{13}|^{2}$)~\cite{Boz1,Boz2}.
In Eq. (\ref{diagonal}), the  radiatively corrected (3$\times$3) dimensional Higgs mass--squared matrix
has been calculated using the effective potential method, 
by taking into account the dominant top-stop as well as the 
bottom-sbottom effects, and the elements of the  Higgs mass--squared matrix can be 
found in \cite{Boz2}.

The other contributions to two-loop eEDM come from the  Higgs boson
couplings  to top-bottom quarks~\cite{Pilaftsis3},
and can be expressed in the following form:
\begin{eqnarray}
\label{topbottom}
\bigg(\frac {d_e}{e}\bigg)^{2-loop}_{H_i- q } &=& 
b_0 \frac{m_e}{M_W^2} \alpha^2  \sum_{i=1}^3
\Bigg \{ - t_\beta {\cal R}_{i3}\bigg[
G_{H_i b t}^{1 S}  {\cal R}_{i1} +G_{H_i b t}^{2 S}  {\cal R}_{i2} + 
G_{H_i bt}^{3S}  {\cal R}_{i3}\bigg] \nonumber\\
&+&  
 \frac { {\cal R}_{i1}} {c_{\beta}}
\bigg[ G_{H_i bt}^{1 P}  {\cal R}_{i1} +G_{H_i bt}^{2 P}  {\cal R}_{i2} +
G_{H_i bt}^{3 P}  {\cal
R}_{i3} \bigg]\Bigg \} 
\end{eqnarray}
Here,  ${\cal R}_{ij}$ is defined in Eq. ($\ref{diagonal}$),  
and $b_0=-3 /8 \pi^2 s_W^2$, whereas 
\begin{eqnarray}
G_{H_i b t}^{1, 2 S(P)}&=& \frac{Q_b^2}{ \ c_\beta}\, \mbox{Re} [ g_{1, 2, bb}^{S(P)} ]
f(g)(m_b^2, m_{{H}_{i}}^2) 
+  \frac{Q_t^2}{s_{\beta}} \, \mbox{Re} [ g_{1,2, tt}^{S(P)} ]  f (g)(m_t^2,
m_{{H}_{i}}^2 )~,\nonumber\\ 
G_{H_i b t}^{ 3 S(P)}&=& Q_b^2 \, \mbox{Re} [ g_{3,bb}^{S(P)} ] f(g)(m_b^2, m_{{H}_{i}}^2 ) 
+Q_t^2 \,  \mbox{Re}[ g_{3,tt}^{S(P)} ] f(g)(m_t^2, m_{{H}_{i}}^2). 
\label{gtb}
\end{eqnarray}
combine the loop functions  $f(g)(m_q^2,
m_{{H}_{i}}^2)$ with the elements of the coupling coefficients \,\,\,
$g_{1,2,3, bb(tt)}^{S(P)}$~\cite{Pilaftsis3}, for  $Q_{t(b)}=\frac{2}{3}(-\frac{1}{3})$. 

Finally, the  Higgs boson
couplings  to  charginos are given by~\cite{Pilaftsis3}:
\begin{eqnarray}
\label{chargino2}
\bigg(\frac {d_e}{e}\bigg)^{2-loop}_{ H_i- \chi_j^{+}} &=& 
-c_0 \frac{m_e \alpha^2} {M_W } \Bigg \{ 
\sum_{i=1}^3  \sum_{j,k=1,2} \frac{1} { {m_{\chi_j^{+}}}} \bigg[-t_{\beta}
{\cal R}_{i3}\,\big(  {\cal C}_{kj}^{1}   {\cal R}_{i1}+  {\cal C}_{kj}^{2}
{\cal R}_{i2}\nonumber\\&+&   
{\cal C}_{kj}^{3}   {\cal R}_{i3}\big)
+  \frac{ {\cal R}_{i1}}{c_\beta} \, 
  \big(  {{\cal C}^{\prime}}^{1}_{kj}   {\cal R}_{i1}+  {{\cal
C}^{\prime}}^{2}_{kj}   {\cal R}_{i2}
+ {{\cal C}^{\prime}}^{3}_{kj}   {\cal R}_{i3}\big)\bigg]\Bigg \}~,
\end{eqnarray}
where $c_0= b_0/3 \sqrt{2}$, and ${\cal C}_{kj}$ 
terms represent  different  combinations of  the 
$2\times 2$ unitary matrices $({\cal U}  {\cal V})$, which
diagonalize the chargino mass matrix, multiplied by the 
loop functions  $f(g)(m_{{ \chi}_j^{+}}^{2}, m_{{H}_{i}}^2)$~\cite{Pilaftsis3}.

\section{Numerical Analysis}
In the following, we will perform a numerical study
to determine the possible 
constraints on $\tan\beta$, $|\mu|$
and the physical phases of the model.
In doing this, we use the present experimental upper bound of the 
electron EDM~\cite{Commins,Abdullah}:
\begin{eqnarray}
d_e < 4.3  \times 10^{-27} \, \mbox{e.cm}~,
\end{eqnarray}  
and impose simultaneously the LEP lower limit on the  Higgs mass:
$m_{h_{3}}\simgt 115~\mbox{GeV}$ (and correspondingly $\tan\beta \simgt 3.5
$) \cite{LEP1,LEP2}, and all lower bounds on the sparticle masses from direct
searches.  We would like to note that here
we are performing the worst case analysis, that is,  we are looking for a Higgs boson
which is well inside the existing experimental bounds. Although one
can analyze SUSY models with a much lower bound~\cite{LEP2},
it is important to look for a Higgs boson which has left no trace in LEP data
irrespective of the model adopted, SM, or SUSY, or any other
extension of SM.

In our analysis, we will particularly concentrate  on the
lightest Higgs boson,  whose CP--odd component as well as its mass are of
prime importance in direct Higgs boson searches at 
high-energy colliders~\cite{Carena2}. In fact, we  use the lightest Higgs boson  as an experimental probe to 
analyze the dependence of its mass, and  CP-odd component 
on the CP-violating phases of the model, in the parameter space allowed  by the EDM constraints. 

A convenient way to observe the effects of the EDM constraints, is via the dimensionless quantity:
\begin{eqnarray}
\mbox{eEDM}=  \frac{{(d_e/e)}^{th}}{{(d_e/e)}^{exp}}~,
\end{eqnarray}
which measures the fractional enhancement or suppression
of the eEDM with respect to its  experimental value. 

Being a reflecting property of the model, 
all the soft masses are expressed in terms of the $\mu$
parameter, and since the $\mu$
parameter is already stabilized to the weak scale  as a consequence of
naturalness, all  dimensionless parameters 
are expected to be of O(1). 
One thus notes that when all k parameters are
of  O(1), all squark soft masses,  trilinear couplings, 
and the gaugino masses scale in exact proportion with the $\mu$
parameter.
\begin{figure}[htb]
\vspace*{0.1truein}
\vspace*{10pt}
\centerline{\epsfig{file=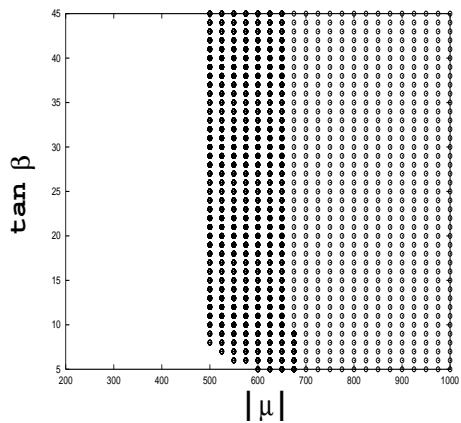,height=2.2in,width=2.4in}}
\vspace*{0.1truein}
\caption{ The interdependence of $\tan\beta$ on  $|\mu|$, when all the phases
are changing from 0 to $\pi$,  $\tan\beta$ from 5 to 45, and  $|\mu|$
from 200 to 1000 $\mbox{GeV}$.}
\label{fig1}
\end{figure}
\begin{figure}
\vspace*{0.1truein}
\vspace*{10pt}
\centerline{\epsfig{file=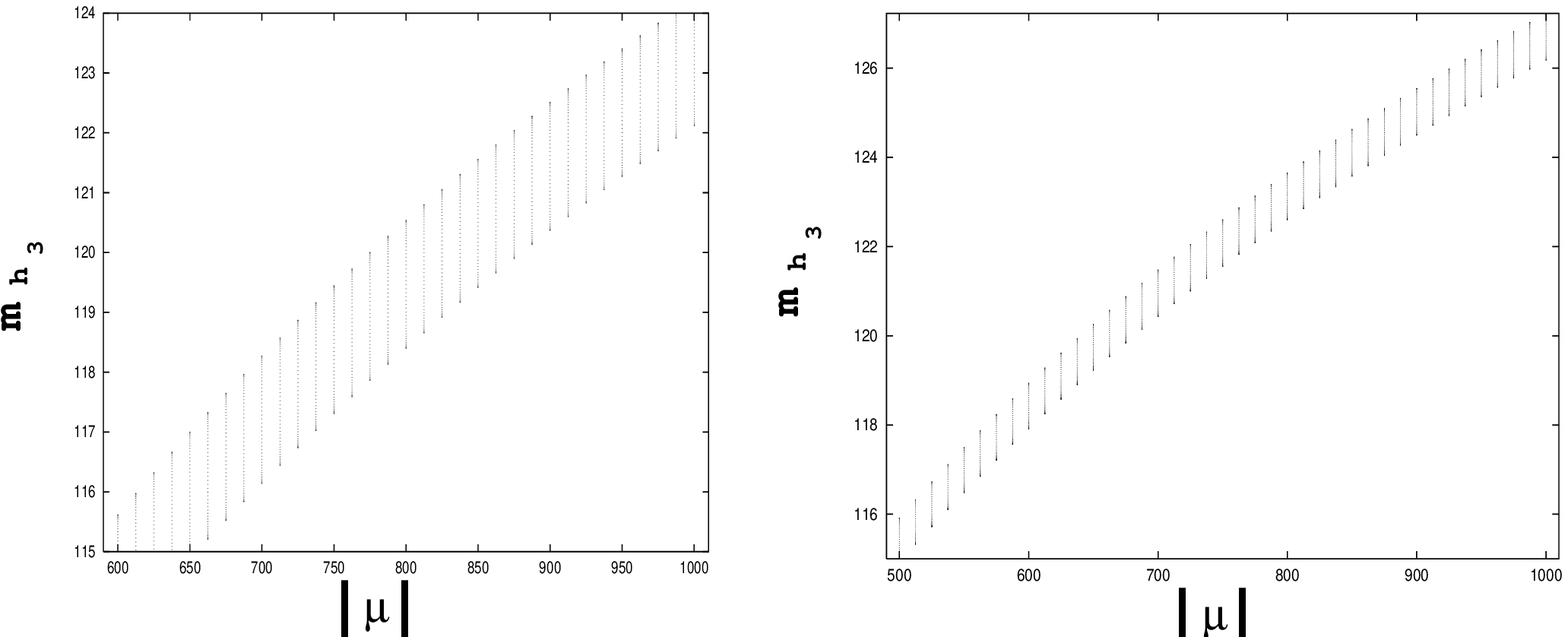,height=2.2in,width=5in }}
\vspace*{0.2truein}
\caption{The dependence of the lightest Higgs boson mass ($m_{h_{3}}$) on $|\mu|$ when $\tan\beta=5$ (left
panel), and $\tan\beta=45$ (right panel), when all the phases
are changing from 0 to $\pi$, and  $|\mu|$ from 200 to 1000 $\mbox{GeV}$.}
\label{fig2}
\end{figure}

Our starting point is the general case
for which we vary: $(i)$ the phases of stop, sbottom, selectron
trilinear couplings  ($\varphi_{A_t}$,  $\varphi_{A_b}$,  $\varphi_{A_e}$),
and the phases of the hypercharge gaugino and SU(2) gaugino
masses ($\varphi_{1}$,  $\varphi_{2}$) from  0 to $\pi$,
$(ii)$ $\tan\beta$ from 5 to 45, and  $(iii)$ $|\mu|$
from 200 to 1000 $\mbox{GeV}$.
Then, in Fig. 1,  we show the dependence of $\tan\beta$ on $|\mu|$,
As Fig. 1 suggests,  the lower allowed bound on $\mu$,
being $500 ~{\rm GeV}$  for  $\tan\beta \simgt 7$,
is pushed to $600 ~{\rm GeV}$ at  $\tan\beta \simlt 7$.
For $|\mu| \simgt  600 ~{\rm GeV}$, all values of
$|\mu|$ and $\tan\beta$ are allowed in the full domain. 
One notes that, the region of the parameter space for which  
$|\mu|$ $\simlt$  $500~\mbox{GeV}$ 
is completely forbidden by the existing constraints on the model, in particular 
by the experimental constraint on the lightest Higgs 
boson mass~\cite{LEP1}, 
as will be indicated in Fig. 2.

To have a better understanding of the effects of the LEP constraint on the
full parameter space, we choose two values of $\tan\beta$,
$\tan\beta=5$, and $\tan\beta=45$ representing the low and high $\tan\beta$
regimes, respectively, and  show the dependence of the lightest Higgs boson mass
($m_{h_{3}}$) on $|\mu|$
in Fig. 2, when all the phases change from 0 to $\pi$. 

As can be seen  from Fig. 2 that  
those portions of the parameter space,  corresponding to   $|\mu| \simlt 600
~{\rm GeV}$ for  $\tan\beta= 5$, and  $|\mu| \simlt 500
~{\rm GeV}$ for  $\tan\beta= 45$, are  disallowed by the experimental 
constraint on the lightest Higgs boson mass 
which requires  $m_{h_{3}}\simgt 115~\mbox{GeV}$ \cite{LEP1}. 
A comparative look at both Fig 1, and Fig. 2 suggests that
the LEP bound puts important constraints on the parameter space of 
the model under concern. In general, when all k parameters are naturally of the order of 1, in size,
$|\mu|$ can not take values below  $500 \mbox{GeV}$ for $\tan\beta
\simgt 7$, and $600 \mbox{GeV}$ for $ 5 \simlt \tan\beta \simlt 7$.
Therefore, constraints from EDM, even if $\tan\beta$
is large, do not exclude a large portion of the parameter space,  because of
the fact that sparticles are heavy enough to have negligible contributions
to the one and two-loop EDMs.
\begin{figure}[htb]
\vspace*{0.1truein}
\vspace*{10pt}
\centerline{\epsfig{file=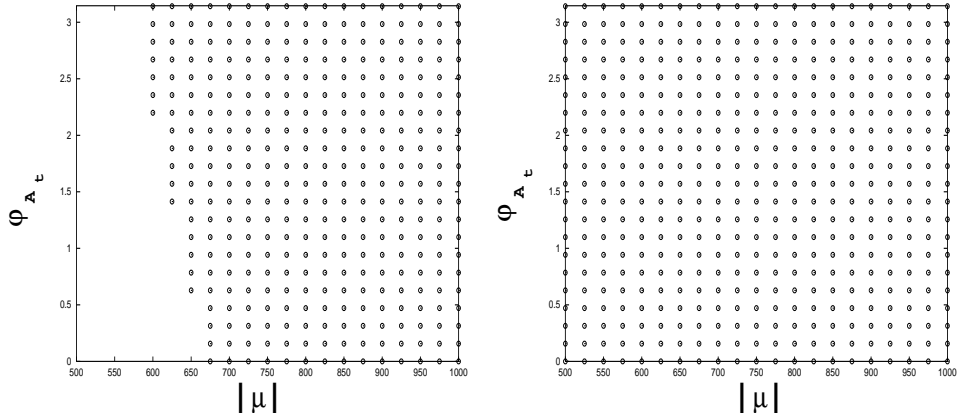,height=2.2in,width=5in }}
\vspace*{0.1truein}
\caption{ The dependence of $\varphi_{A_{t}}$  on $|\mu|$, for 
$\tan\beta=5$~(left panel), and $\tan\beta=45$~(right panel), when all the
phases are changing from 0 to $\pi$.}
\label{fig3}
\end{figure}

The analyses of Fig. 1 and Fig. 2 give a general idea of
the allowed parameter domain in the   $\tan\beta-|\mu|$ plane,
when all the phases vary in the full range.
With this input in mind,  to study  the
possible constraints on the physical phases,
we first explore the dependence of  
$\varphi_{{A}_{t}}$ on $|\mu|$ for low and high values of $\tan\beta$, in
Fig.~3.

In Fig.~3, we  show the dependence of  $\varphi_{A_{t}}$ on  $|\mu|$  for
$\tan\beta=5$~(left panel), and $\tan\beta=45$~(right panel),
when all the phases are changing from 0 to $\pi$.
As suggested by the left panel of Fig. 3, at low
values of  $\tan\beta$, where the two--loop eEDM contributions are negligible, 
certain values of $|\mu|$
and $\varphi_{A_t}$ are excluded. For instance,
at $\tan\beta=5$ and $|\mu|=650\ {\rm GeV}$,
the portion  of the parameter space
for which $\varphi_{A_t} \simlt \pi/5 $ is forbidden.
The allowed range of $\varphi_{A_t}$
gets norrower until  $\varphi_{A_t} \sim 7 \pi/10 $,
as $\mu$ changes  from $650$ to  $600\ {\rm GeV}$. 
For lower values of $\mu$ there is
no allowed domain at all. 
On the other hand, at $\tan\beta=45$ 
all points in  $\varphi_{A_t}-|\mu|$ plane are allowed for $\mu\simgt 500\ {\rm GeV}$, as shown in the right
panel of Fig.~3.

A comparative look at both panels of Fig. 3 suggests that, the allowed range of
$\varphi_{A_t}$ and ($\tan\beta$, $|\mu|$)
become  strongly correlated  particularly, at low
$\tan\beta$. As the allowed range of  $\varphi_{A_t}$ gradually widens
(from $7 \pi/10$  to $\pi/5$),
the lower bound on $|\mu|$ changes from  $600 {\rm GeV}$ to  $675 {\rm GeV}$,  in
the low $\tan\beta$ regime (where the main contribution comes 
from the one-loop eEDM).
This constraint on 
$\varphi_{A_t}-|\mu|$ domain is lifted in the high $\tan\beta$ regime, due to the cancellations between
one-- and two--loop EDMs.
\begin{figure}[htb]
\vspace*{0.1truein}
\vspace*{10pt}
\centerline{\epsfig{file=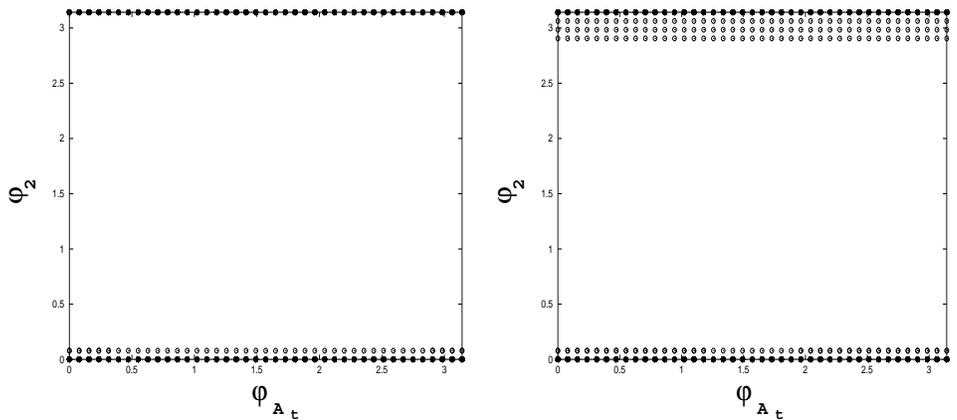,height=2.2in,width=5in }}
\vspace*{0.1truein}
\caption{ The dependence of 
$\varphi_{2}$ on  $\varphi_{A_{t}}$,
for $\mu=700 {\rm GeV}$ (left panel), and $\mu=1000 {\rm GeV}$ (right panel),
at $\tan\beta=5$, when all the phases changing from 0 to $\pi$.}
\label{fig4}
\end{figure}

To understand the interdependence of
the phase of the  $SU(2)$ gaugino mass ($\varphi_{2}$)
on $\varphi_{A_{t}}$, and to explore the possible constraints
on $\varphi_{2}$, we first focus on the low-$\tan\beta$
regime, where   $\varphi_{A_{t}}$ is quite sensitive to the lower bound on
$|\mu|$. We choose two particular values
of $|\mu|$ at $\tan\beta=5$, and
in Fig. 4, we show the variation of $\varphi_{2}$
with  $\varphi_{A_{t}}$  for  $|\mu|=700~\mbox{GeV}$~(left panel),
and  $|\mu|=1000~\mbox{GeV}$~(right panel), when all the phases
changing from 0 to $\pi$, as  in Fig. 3.
The left panel of the Fig. 4 suggests that
as  $\varphi_{A_{t}}$ varies  in its
full range at   $|\mu|=700~\mbox{GeV}$ (as has been suggested also by Fig. 3),
$\varphi_2$ remains  in the vicinity of
$0 \simlt \varphi_2 \simlt \pi/20$. For higher values of $\varphi_2$,
no solutions can be found in the parameter space until $\varphi_2=\pi$.
One notes that, $\varphi_{A_{t}}-\varphi_2$ domain is the similar,
for $|\mu| \simlt 700~\mbox{GeV}$,
except for the lower allowed bound of $\varphi_{A_{t}}$ which changes from $ 7 \pi/10$
to $\pi/5$ in the  $ 600 \simlt |\mu| \simlt 700~\mbox{GeV}$ interval.

On the other hand,  for higher values of $|\mu|$,
for instance  at   $|\mu|=1000~\mbox{GeV}$ (right panel of Fig. 4),
the allowed domain of $\varphi_{2}-\varphi_{A_{t}}$  slightly
widens, however,
$\varphi_{2}$ still remains
in the vicinity of CP-conserving points.
The sensitivity of  $\varphi_{2}$
becomes stronger in the  high-$\tan\beta$
regime, where  the two-loop eEDM
effects dominate~\cite{Chang,Pilaftsis3}.
\begin{figure}[htb]
\vspace*{0.1truein}
\vspace*{10pt}
\centerline{\epsfig{file=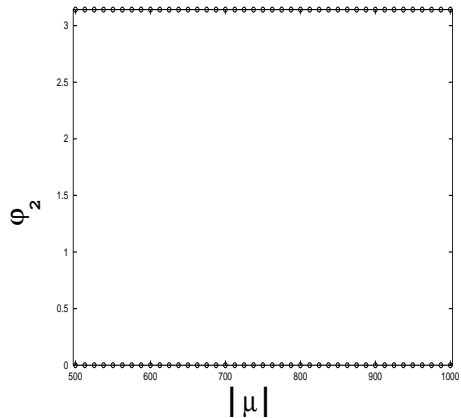,height=2.2in,width=2.4in}}
\vspace*{0.1truein}
\caption{The dependence of 
$\varphi_{2}$ on $|\mu|$  at $\tan\beta=45$,
when all the phases changing from 0 to $\pi$.}
\label{fig5}
\end{figure}

For instance, in Fig.~\ref{fig5},
we show the dependence of  $\varphi_{2}$
on $|\mu|$  at $\tan\beta$=45, when all the phases change from 0 to $\pi$,
like in all previous cases. It  can be seen from Fig.~\ref{fig5} that
as $|\mu|$ changes from 500 $\mbox{GeV}$ to  1000 $\mbox{GeV}$,
the phase of the SU(2) gaugino  mass is imprisoned to lie at a CP conserving
point, for all values of  $\varphi_{A_{t}}$, in the high-$\tan\beta$ regime.

A comparative analysis of Fig.~\ref{fig4} and Fig.~\ref{fig5}
suggests that the CP violating phase from the char\-gino sector
is required to be in close vicinity of CP conserving points
in the low-$\tan\beta$ regime depending on $|\mu|$,
whereas it  remains stuck completely to CP conserving points
in the  high-$\tan\beta$, 
for all values of $|\mu|$ 
$\simgt 500~\mbox{GeV}$.
One notes that  
when all phases are changing from 0 to $\pi$,
all points in the $\varphi_{1}-|\mu|$ 
plane are allowed in the $600 \simlt|\mu| \simlt 1000~\mbox{GeV}$
interval, at low- $\tan\beta$ regime ($\tan\beta=5$).
The parameter domain of $\varphi_{1}-|\mu|$ plane is similar in the high-$\tan\beta$ regime
($\tan\beta=45$), but the lower allowed bound of $|\mu|$ 
becomes  $500~\mbox{GeV}$ for this case. 

Since  the  phase of the SU(2) gaugino mass  
$\varphi_{2}$, is sensitive to $\tan\beta$, and $|\mu|$
parameters,  and  it is 
required to  be in close   vicinity of CP conserving points in general, 
we will take it  it to be at $\varphi_{2}=\pi$,  
for the rest of the analysis, which  we  focus on 
various parameter planes for analyzing the dependence of eEDM
on the  phases of the trilinear couplings.  In doing this,  we will vary  each of the physical phases 
in the full $[0,\pi]$ range, setting all the others at the  maximal CP violation point.

\subsection{$|eEDM|$ versus $\varphi_{A_{t}}$}

We show the dependence of $|eEDM|$ on $\varphi_{{A}_{t}}$ in Fig. 6,
at  $\tan\beta=5$ (left panel), and $\tan\beta=45$ (right  panel), when 
$\varphi_{1}=\varphi_{A_{b}}= \varphi_{A_{e}}= \varphi_{1,b,e}= \pi/2$,
and  $\varphi_{A_{t}}$ changes in the  [0, $\pi$] interval.
\begin{figure}[htb]
\vspace*{0.1truein}
\vspace*{10pt}
\centerline{\epsfig{file=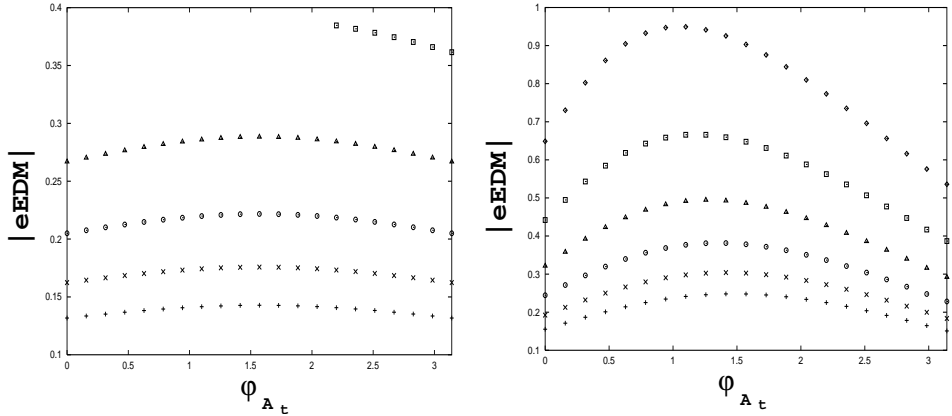,height=2.2in,width=5in }}
\vspace*{0.1truein}
\caption{The dependence of $|eEDM|$  on $\varphi_{A_{t}}$
at $\tan\beta=5$ (left panel), and $\tan\beta=45$ (right panel),
when  $\varphi_{1,b,e}= \pi/2$. In both panels, those  portions of the parameter space 
presented by the curves  $``\Box''$, $``\triangle''$,  $''\circ''$,  $''\times''$, 
belong to $ 600, \,  700,\,  800,\,  900,\,  1000  \mbox{GeV}$
values of $|\mu|$, respectively, whereas   $''\diamond''$ corresponds to $|\mu|=
500\mbox{GeV}$ in the right panel.}
\label{fig6}
\end{figure}

In both panels of Fig. 6, the various curves 
shown by $``\Box''$, $``\triangle''$,  $''\circ''$,  $''\times''$,
$''+''$ belong to $ 600, \,  700,\,  800,\,  900,\,  1000  \mbox{GeV}$
values of $|\mu|$, respectively, whereas  $''\diamond''$ corresponds 
to $|\mu|=500\mbox{GeV}$ in the right panel.
As the left panel of the Figure suggests,
in the low-$\tan\beta$ regime (where the two-loop EDM contributions are
small), $|eEDM|$   maximally extends to  $\sim 0.4$ at  $\mu= 600 \mbox{GeV}$
($''\Box''$), and at $ \varphi_{{A}_{t}}\simgt 7 \pi/10$.
The remaining portion of the parameter space ($\varphi_{{A}_{t}}\simlt 7
\pi/10$) is discarded by the existing constraints on the model (see, Fig. 3).  
On the other hand,  as $\mu$ gets larger  values,  $|eEDM|$
decreases.  For instance, when $|\mu|= 1000 \mbox{GeV}$ \, ($''+''$), the maximal value of  $|eEDM|$
does not exceed $\sim 0.15$ in the low-$\tan\beta$ regime (left panel).
For the  high-$\tan\beta$ regime (right panel),
the upper bound of the $|eEDM|$ increases, 
whereas  the allowed range of $|\mu|$ widens, as compared to  $\tan\beta=5$ case. 
For instance, as $\varphi_{{A}_{t}}$  ranges in the full  $[0, \pi] $
interval, the  maximal value of  $|eEDM|$  occurs  at $|\mu|=500 \mbox{GeV}$
($''\diamond''$) when   $\tan\beta=45$. 
A comparative look at both panels of Fig. 6  suggests  that 
$|eEDM|$
grows with $\tan\beta$, and 
it  approaches to the upper bound,
particularly in  the  high-$\tan\beta$ regime, since the two-loop 
EDM contribution grows with $\tan\beta$.
\begin{figure}[htb]
\vspace*{0.1truein}
\vspace*{10pt}
\centerline{\epsfig{file=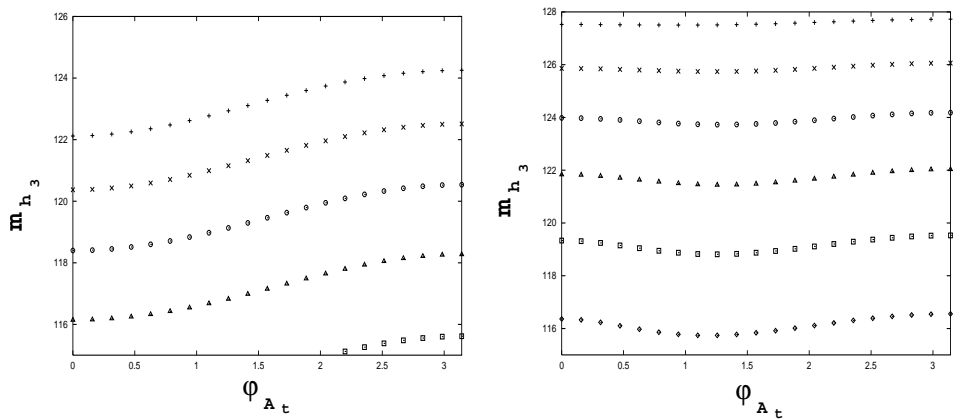,height=2.2in,width=5in }}
\vspace*{0.1truein}
\caption{The dependence of mass of the lightest Higgs boson ($m_{h_{3}}$) on 
$\varphi_{A_{t}}$ at  $\tan\beta=5$ (left panel), and $\tan\beta=45$ (right panel),
when  $\varphi_{1,b,e}= \pi/2$. In both panels,  $``\Box''$, $``\triangle''$,  $''\circ''$,  $''\times''$, 
represent  $ 600, \,  700,\,  800,\,  900,\,  1000 \,  \mbox{GeV}$
values of $|\mu|$, respectively, whereas   $''\diamond''$ corresponds to 
$|\mu|=500\mbox{GeV}$ in the right panel.}
\label{fig7}
\end{figure}

In Fig.~7, we show the  variation of the
mass of the lightest Higgs boson
($m_{h_{3}}$) with $\varphi_{A_t}$,
at $\tan\beta=5$ (left panel), and $\tan\beta=45$
(right panel), 
when  $\varphi_{1,b,e}= \pi/2$.
As in  Fig. 6, in both panels of Fig. 7,
the curves shown by $``\Box''$, $``\triangle''$,  $''\circ''$,  $''\times''$,
$''+''$, present  600,  700,  800,  900,  1000~$\mbox{GeV}$,
values of $|\mu|$, respectively,  whereas $''\diamond''$ belongs to 
$|\mu|=500\mbox{GeV}$ in the right panel. It can  be seen from the left panel that,
being  $\simlt 116 \mbox{GeV}$  at the lower
bound~($|\mu|= 600 \mbox{GeV}$, \, $''\Box''$),
$m_{h_{3}}$  maximally extends
to  $\sim 124 \mbox{GeV}$ when  $|\mu|= 1000 \mbox{GeV}$ \, ($''+''$), in the
low $\tan\beta$ regime. 
Therefore, as $|\mu|$
gets larger values,  $m_{h_{3}}$ increases,
which  remains true also in the
high-$\tan\beta$ regime (right panel).
A comparative look at both  panels of 
Fig. 7 suggests that $m_{h_{3}}$ is much more
sensitive to the variations in  $\varphi_{A_{t}}$ at
$\tan\beta=5$~(left panel), than that of $\tan\beta=45$  (right panel),
since the radiative corrections depend strongly on the stop splitting 
at low-$\tan\beta$ regime. On the other hand,  the dependence of $m_{h_{3}}$
on  $\varphi_{A_{t}}$  weakens in passing from the low $\tan\beta$
regime to higher,  since  the radiative corrections to  
$m_{h_{3}}$ which are sensitive to the variations in $\varphi_{A_{t}}$ are suppresed in the 
high-$\tan\beta$ regime \cite{Boz1}.
One notes that  this is the region of the parameter space 
in which $0.1 \simlt |eEDM|  \simlt 0.4 $ for $\tan\beta=5$, and 
$0.1   \simlt  |eEDM| \simlt 0.98$ for $\tan\beta=45$, as has been suggested
by   Fig.~6.
\begin{figure}[htb]
\vspace*{0.1truein}
\vspace*{10pt}
\centerline{\epsfig{file=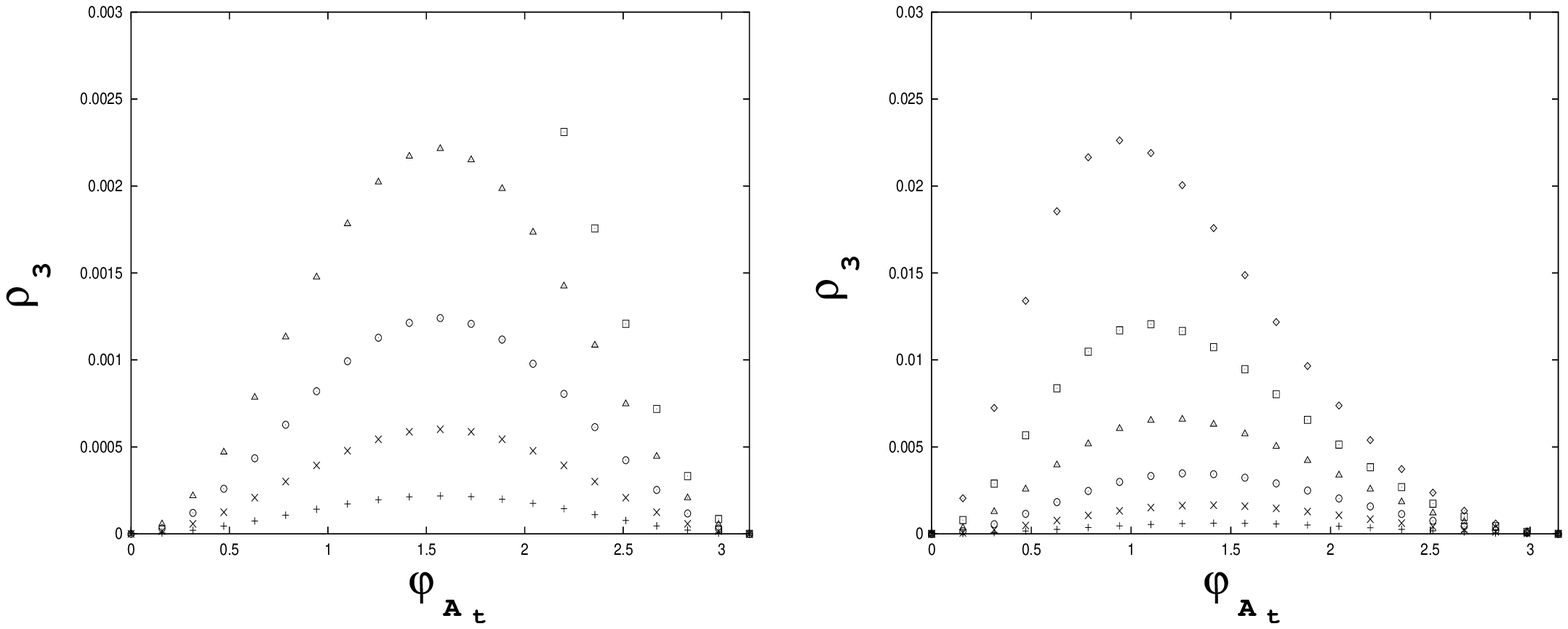,height=2.2in,width=5in }}
\vspace*{0.1truein}
\caption{The dependence of the CP-odd component $(\rho_3)$ of the lightest Higgs boson on  $\varphi_{A_{t}}$
at $\tan\beta=5$ (left panel), and $\tan\beta=45$~(right panel),
when  $\varphi_{1,b,e}= \pi/2$, and    $|\mu|$= 600 \ ($''\Box''$),  700 \
($''\triangle''$), 800 \ ($''\circ''$), 900\ ($''\times''$),  1000 \
($''+''$)~$\mbox{GeV}$~(in both panels), and
$|\mu|=500\mbox{GeV}$~($''\diamond''$,  right panel)}
\label{fig8}
\end{figure}

In Fig.~8,  we show the  variation of the  CP-odd component ($\rho_3$)
of  the lightest Higgs boson
with $\varphi_{A_t}$, for  $\tan\beta=5$ (left panel), and $\tan\beta=45$
(right panel), at   $\varphi_{1,b,e}= \pi/2$,  
when  $|\mu|$= 600 \ ($''\Box''$),  700 \
($''\triangle''$), 800 \ ($''\circ''$), 900\ ($''\times''$),  1000 \
($''+''$)  $\mbox{GeV}$ (in both panels), and     
$|\mu|=500\mbox{GeV}$\,  ($''\diamond''$,  right panel). 
As both panels of the  Figure suggest,
higher the $|\mu|$, smaller the $\rho_3$.
Such kind of $\rho_3-|\mu|$ interdependence is expected, since we
particularly focus on the region of the parameter space for which the
scale dependence is sufficiently suppressed 
( $1000 \simlt Q  \simlt 1200$) \cite{Boz2}, and 
we set, for convenience,  ${\rm Q}=1200 {\rm GeV}$.
The properties of various renormalization scales, changing from top mass
to $\mbox{TeV}$ scale, and particularly their influences on
$\rho_3-\mu$ plane has been studied in Ref.~[28].

The dependence of $m_{h_{3}}$, and $\rho_3$
on $\varphi_{A_{t}}$  in the parameter space allowed by  the EDM constraints
shows that, being  quite sensitive to the variations in $\varphi_{A_{t}}$,
 $m_{h_{3}}$  maximally reaches to $ \sim 128 \mbox{GeV}$  when $\mu=1000\mbox{GeV}$, and  
$\tan\beta=45$. On the other hand,  $\rho_3$ remains below $\sim 0.3\%$ though 
it grows by more than an order of magnitude as $\tan\beta$ changes from 5 to 45. This is 
similar to constraints found for MSSM 
Higgs sector \cite{Demir1}. 

\subsection{$|eEDM|$ versus $\varphi_{A_{e}}$}
In Fig. 9, we show the variation of $|eEDM|$ with the selectron trilinear
coupling ($\varphi_{A_{e}}$)
at   $\tan\beta=5$ (left panel), and $\tan\beta=45$ (right panel),
when  $\varphi_{1}=\varphi_{A_{t}}=\varphi_{A_{b}}=\varphi_{1, t,b}=\pi/2$, and
$\varphi_{A_{e}}$  changes in the [0, $\pi$] interval.
\begin{figure}[htb]
\vspace*{0.1truein}
\vspace*{10pt}
\centerline{\epsfig{file=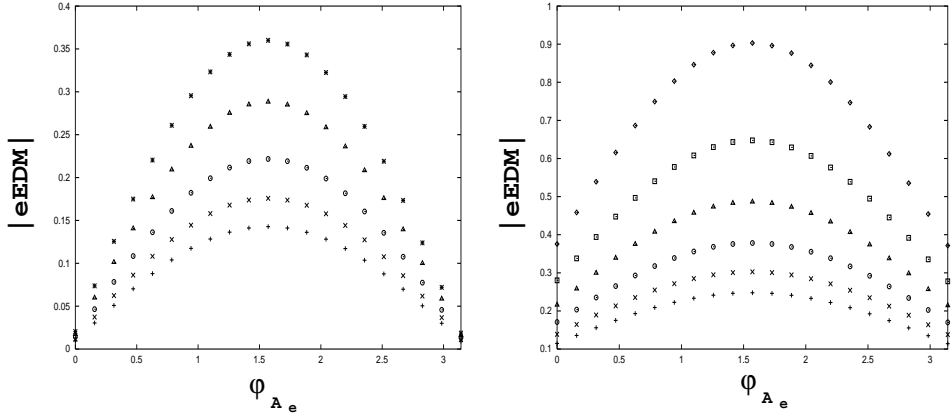,height=2.2in,width=5in }}
\vspace*{0.1truein}
\caption{The dependence of $|eEDM|$ on $\varphi_{A_{e}}$
at  $\tan\beta=5$ (left panel), and $\tan\beta=45$ (right panel),
when $\varphi_{1, t,b}=\pi/2$. In the figure,
$500, \, 600, \ 700, \,  800, \, 900, \, 1000 \mbox{GeV}$
values of $|\mu|$ are presented by the curves
$''\Box''$,  $''\diamond''$,
$''\triangle''$,  $''\circ''$,  $''\times''$,  $''+''$,
whereas  $|\mu|= 625 \mbox{GeV}$ ($''*''$)
is the lower allowed bound at $\tan\beta=5$ (left panel)}
\label{fig9}
\end{figure}

In the figure,  $500, \, 600, \  700, \,  800, \, 900, \, 1000~\mbox{GeV}$,
values of $|\mu|$ are presented by the curves
$''\Box''$,  $''\diamond''$,
$''\triangle''$    ,$''\circ''$,  $''\times''$,  $''+''$,
whereas    $''*''$ (left panel),
corresponds to $|\mu|= 625 \mbox{GeV}$.
One notes that,  at $\tan\beta=5$ (left panel),  the region of the parameter space for which
 $|\mu| \simlt  625 \mbox{GeV}$ is forbidden by the
existing constraints on the model. Indeed, remembering the  $\varphi_{A_{t}} -|\mu|$
domain in  Fig. 3 for instance, it can be  seen that 
the lower allowed bound on $|\mu|$ is pushed from $600$ to  $625 \mbox{GeV}$ for  $\varphi_{A_{t}} \simgt
\pi/2$ at $\tan\beta=5$. 
Such a constraint is lifted in the high-$\tan\beta$
regime, and all points in the  $\varphi_{A_{t}}-\varphi_{{A}_{e}}$
are allowed for  $|\mu| \simgt 500 \mbox{GeV}$.
As both panels of the Figure suggest,  
the  variation  of $\varphi_{A_{e}}$  around  $\varphi_{A_{e}}=\pi/2$
differs from those at   $\varphi_{A_{e}}=0$, and  $\varphi_{A_{e}}=\pi$.
That is, increasing  with $\varphi_{A_{e}}$ in the $[0,\pi/2]$ interval,
the  maximal value of  $|eEDM|$ occurs at  $\varphi_{A_{e}}=\pi/2$. 
Then,  it gradually decreases  in the   
$[\pi/2,\pi]$ interval.

\subsection{$|eEDM|$ versus $(\tan\beta, |\mu|)$}
Finally, in Fig. 10, we have shown the dependence of  $|EDM|$
on  $|\mu|$, when the physical phases of the 
model are chosen as: $\varphi_{1}=\varphi_{A_{t}}=
\varphi_{A_{b}}=\varphi_{A_{e}}=\varphi_{1tbe}=\pi/2$,
whereas  $\varphi_{2}=\pi$, like all the previous cases.
\begin{figure}[htb]
\vspace*{0.1truein}
\vspace*{10pt}
\centerline{\epsfig{file=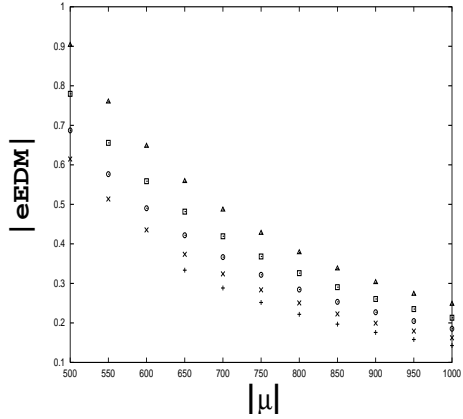,height=2.2in,width=2.4in }}
\vspace*{0.1truein}
\caption{The dependence  of $|eEDM|$ on $|\mu|$,
at $\tan\beta= 5 \, (''+'') $, $15 \, (''\times'')$,  
$ 25 \, (''\circ'')$,  $ 35\,  (''\Box'')$, $ 45 \, (''\triangle'') $,
when  $\varphi_{1tbe}=\pi/2$.}
\label{fig10}
\end{figure}

In Fig. 10, we have chosen various values of $\tan\beta$,
which are represented by  the curves:  $\tan\beta$
= 5 \,($''+''$),
\,\,   $\tan\beta= 15$ \, ($''\times''$), \,  $\tan\beta= 25$ \,($''\circ''$),
$\tan\beta= 35$ \,($''\Box''$),  $\tan\beta= 45$ \,($''\triangle''$).
As  Fig.~10 suggests, when $\varphi_{1tbe} \leadsto \pi/2$,
being  $\sim 0.35$ for $\tan\beta=5$,
the  maximal value of $|eEDM|$  reaches beyond $\sim 0.9$ for  $\tan\beta=45$,
when $\varphi_{1tbe} \leadsto \pi/2$.
Therefore, as has been mentioned in the previous cases (for instance,
Fig.~5), the general tendency of  $|eEDM|$ is such that
it grows with  $\tan\beta$. However,
the dependence of $|eEDM|$ on $|\mu|$  differs from that
of $\tan\beta$ in the sense that $|eEDM|$ decreases as $|\mu|$ gets
higher values. 

In Fig. 11, which  supplements Fig. 10, the  variation of
$|eEDM|$ with $\tan\beta$ is shown, when $\varphi_{1tbe} \leadsto \pi/2$.
Here, the curves shown by  $ ''+''$, \,  $''\circ''$, \,  $''\Box''$,
\,$''\triangle''$,\, $''\diamond''$
represent  $\mu$ = 500,\,
600, \,700,  \,800,  \, 900,  and   1000 \,$\mbox{GeV}$.
values of $|\mu|$. 
As the Figure suggests, when  $\tan\beta=45$, $|eEDM|$  occurs  at $\sim
0.2$,  at $\mu$= 1000 $\mbox{GeV}$. Increasing gradually, it reaches
far below  $\sim 0.5$ at  $\mu= 700 \mbox{GeV}$, and  finally  when $\mu= 500
\mbox{GeV}$,   $|eEDM|$ approches to the upper bound.
\begin{figure}[htb]
\vspace*{0.1truein}
\vspace*{10pt}
\centerline{\epsfig{file=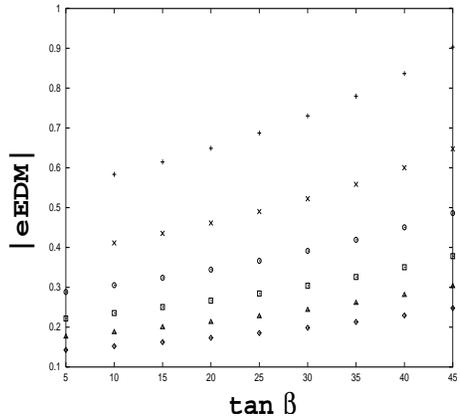,height=2.2in,width=2.4in }}
\vspace*{0.1truein}
\caption{The dependence  of $|eEDM|$ on $\tan\beta$
for various values of $|\mu|$ when $\varphi_{1tbe}$=$\pi/2$. 
Here, $|\mu|$ = 500 \, $(''+'')$,\,  600 \,$(''\times'')$, 
\,700 \,$(''\circ'')$, \,800 \,$(''\Box'')$, \,900
\,$(''\triangle'')$, \, 1000 \, $(''\diamond'')~\mbox{GeV}$.}
\label{fig11}
\end{figure}

Therefore, similar to observations made for \,\,\, the 
former case (Fig. 10), 
one notices that,  
as $|\mu|$ gets larger values,  $|eEDM|$ decreases.
The reason for that is, as has been suggested by Eqs. (\ref {1}-\ref{2}), 
all the soft mass parameters of the model are  expressed in terms of the $\mu$
parameter, and  clearly, as $|\mu|$ gets larger values, the sparticle masses  
increase. 

Taking into account of the fact that eEDM decreases,
as the sparticle masses increase in general,
such kind  of   $|eEDM|-|\mu|$ 
dependence is expected
(higher the $|\mu|$ parameter, heavier the sparticle masses,  and smaller the   SUSY
contributions to the $|eEDM|$).

Before concluding, we would like to note that, although we did not show
explicitely, the branching ratio $B \rightarrow X_s \gamma$ remains within
the bounds. The main reason for having agreement with  $B \rightarrow X_s \gamma$
constraints is that  generally $\mu \simgt 500\mbox{GeV}$, and the
pseudoscalar  and charged Higgs bosons, like the sfermions themselves, are
heavy. Therefore, within this spectrum, non-standard contributions to  $B \rightarrow X_s \gamma$
are suppressed~\cite{Borzumati,KaganNeubert1,DegrassiCarena,DemirOlive,KaganNeubert2}.

\section{Conclusion}

We have analyzed the EDM and Higgs mass constraints on the SUSY model
which solve  the strong CP problem via the dynamical phase of the
gluino mass~\cite{DemirMa1}. The model expresses all 
soft breaking masses in terms of the dynamically--generated
$\mu$ parameter where the dimensionless parameters involved
are naturally of order ${\cal{O}}(1)$ and source the SUSY CP violation.

Our general discussion followed by the numerical estimates 
for various parameter planes show that:
when all the phases are changing from 0 to $\pi$,
$(i)$ $|\mu|$ is forbidden to take values typically below $500~{\rm GeV}$
by the LEP constraint on the lightest Higgs mass.
This is an interesting constraint which has been seen to weaken 
EDM constraints due to the heaviness of the superpartners.
$(ii)$ in contrast to the constrained minimal model, or
unconstrained low energy minimal supersymmetric model,
certain values of the trilinear coupling phases are
disallowed due to both the electric dipole moment and the Higgs mass constraints.
$(iii)$as in the minimal supersymmetric model, in general,
the CP violating phase from the chargino sector 
($\varphi_2$) is  required to be in close vicinity of a 
CP-conserving point. 

\section{Acknowledgements}
\noindent
This work was partially supported by the  Scientific and
Technical Research Council of Turkey (T\"{U}B{\.I}TAK) under the project,
No:TBAG-2002(100T108).

\noindent
The author also thanks Durmu\c{s}
Demir for useful e-mail exchange.

\end{document}